\documentstyle[aps,prb,twocolumn,epsf]{revtex}

\begin{document}
\draft
\title{Ordered phase and scaling in $Z_n$ models
and the three-state antiferromagnetic
Potts model in three dimensions}

\author{Masaki Oshikawa}

\address{
Department of Physics, Tokyo Institute of Technology,
Oh-okayama, Meguro-ku, Tokyo 152-8551, Japan
}

\date{July 20, 1999}

\maketitle

\begin{abstract}
Based on a Renormalization-Group
picture of $Z_n$ symmetric models in three dimensions,
we derive a scaling law for the $Z_n$ order parameter
in the ordered phase.
An existing Monte Carlo calculation
on the three-state antiferromagnetic Potts model, which
has the effective $Z_6$ symmetry,
is shown to be consistent with the proposed scaling law.
It strongly supports the Renormalization-Group picture that there is
a single massive ordered phase, although an
apparently rotationally symmetric region in the intermediate
temperature was observed numerically.
\end{abstract}

\pacs{75.10.Hk, 05.50+q}

\section{Introduction}

The symmetry and the dimensionality
are important factors to determine
the universality class of critical phenomena.
The $O(2)$ symmetry is the simplest among the continuous symmetry,
and statistical models with the
$O(2)$ symmetry has been studied intensively.
A natural question then would be
the effect of the symmetry breaking
from the continuous $O(2)$ to the discrete $Z_n$.
A simple spin model with $Z_n$ symmetry is the $n$-state clock model
with a Hamiltonian
\begin{equation}
	H = - \sum_{\langle j, k \rangle} \cos{(\theta_j - \theta_k)},
\end{equation}
where $\langle j,k \rangle$ runs over nearest neighbors, and
$\theta_j$ takes integral multiples of $2 \pi/n$.
The standard XY model with $O(2)$ symmetry is defined by
the Hamiltonian of the same form; the only difference is that
$\theta$ takes continuous values.

The $Z_n$ symmetry is fundamentally different from $O(2)$
because of its discrete nature.
On the other hand, for large $n$, it is natural to expect 
the $Z_n$ symmetry to have similar effects to
that of the $O(2)$ symmetry.
Understanding these two apparently contradictory aspects
is an interesting problem.
Besides the theoretical motivation, there are some
possible experimental realizations of the effective
$Z_n$ symmetry.
For example, the stacked triangular antiferromagnetic
Ising (STI) model with effective $Z_6$
symmetry may correspond to materials such as\cite{CsMnI3} CsMnI$_3$.

In two dimensions, the phase diagram of the $Z_n$ model
is well understood\cite{Nelson} in the framework
of the renormalization group (RG).
For $n \geq 5$, there is
an intermediate phase between the
low-temperature ordered phase with the spontaneously broken $Z_n$
symmetry and the high-temperature disordered phase.
The intermediate phase is $O(2)$ symmetric and corresponds to
the low-temperature phase of the XY model.

On three dimensional (3D) case, Blankschtein et al.\cite{Blank} in 1984
proposed an RG picture of the $Z_6$ models,
to discuss the STI model.
They suggested that the transition between the ordered and
disordered phases belongs to the (3D) XY universality class,
and that the ordered phase reflects the
symmetry breaking to $Z_6$ in a large enough system.
It means that there is no finite region of
rotationally symmetric phase which is similar to
the ordered phase of the XY model. 
Unfortunately,
their paper is apparently not widely known in the related fields.
It might be partly because their discussion was very brief
and not quite clear.

In fact, there has been a long-standing controversy on the
the three-state antiferromagnetic Potts (AFP) model
on a simple cubic lattice, defined by the Hamiltonian
\begin{equation}
	H = + \sum_{\langle j,k \rangle} \delta_{\sigma_j \sigma_k},
\end{equation}
where $\sigma_j = 0,1,2$ and $\langle j,k \rangle$ runs over
nearest neighbor pairs on a simple cubic lattice.
The order parameter of this model is not evident.
However, previous studies revealed that the low-temperature ordered phase,
which is called as Broken Sublattice Symmetry (BSS) phase\cite{BGJ},
corresponds to a spontaneous breaking of the $Z_6$ symmetry.
Thus the effective symmetry of this model
may be regarded\cite{Ono} as $Z_6$,
although it is not apparent in the model.
It is now widely accepted that there is a phase transition with
critical exponents characterized by the
3D XY universality class\cite{WSK,KO,KS,HWS,Kishi},
at temperature $T_c \sim 1.23$
(we set the Boltzmann constant $k_B=1$.)
On the other hand, according to  numerical calculations,
there appears to be an intermediate phase below $T_c$ and
above the low-temperature phase.
While there have been various proposals\cite{Ono,RL,Sun}
for the intermediate region,
most reliable numerical results at present indicates that
the intermediate region appears to be rotationally symmetric phase
which is similar to the ordered
phase of the 3D XY model\cite{KS,HWS,Kishi}.
However, the ``transition'' between the intermediate region
and the low-temperature phase is not well understood.
According to the suggestion in Ref.~\onlinecite{Blank},
the intermediate ``phase'' would be rather a crossover to the
low-temperature massive phase.

On the other hand, there has been a claim of an
intermediate phase\cite{Mitsubo} also in
the $6$-state clock (6CL) model,
which has the manifest $Z_6$ symmetry.
In a recent detailed numerical study, Miyashita\cite{Miya}
found that the intermediate region appears to have
a rotationally symmetric character, as found in the AFP model.
However, through a careful examination of the system size dependence,
he concluded that it is just a crossover to the
massive low-temperature  phase,
and that the rotationally symmetric XY phase does not exist in the
thermodynamic limit.
His conclusion is consistent with the suggestion in Ref.~\onlinecite{Blank}.

In this article, based on the RG picture,
we derive a scaling law of an order parameter which measures
the effect of symmetry breaking from $O(2)$ to $Z_n$.
We demonstrate that the Monte Carlo results on the
AFP model in Ref.~\onlinecite{HWS} is consistent with the
scaling law, supporting the RG picture with a single phase
transition.

\section{Renormalization-Group Picture}

Since the discussion of the
RG picture in Ref.~\onlinecite{Blank} was rather brief,
it would be worthwhile to present the RG picture here,
with some clarifications and more details.
We also make a straightforward extension to
general integer $n$ from the $n=6$ case.

A generic $Z_n$ symmetric model
may be mapped, in the long-distance limit,
to the following $\Phi^4$-type field theory with the Euclidean action
\begin{equation}
S = \int d^3 x
\left[
	| \partial_{\mu} \Phi |^2 + u |\Phi|^2
		+ g |\Phi|^4
		- \lambda_n ( \Phi^n  + \bar{\Phi}^n )
\right]
\end{equation}
with the complex field $\Phi$ and its conjugate $\bar{\Phi}$.
The $\lambda_n$-term is the lowest order term in $\Phi$
which breaks the symmetry from $O(2)$ to $Z_n$.
The phase transition corresponds to the vanishing of
(the renormalized value of) the parameter $u$.
The temperature $T$ in the $Z_n$ statistical system
roughly corresponds to $u$ as $u \sim T - T_c$ where $T_c$ is
the critical temperature.

In the absence of the symmetry breaking $\lambda_n$,
the transition belongs to the so-called 3D XY universality class.
Its stability under the symmetry breaking to $Z_n$
is determined by the scaling dimension of $\lambda_n$
at the 3D XY fixed point.
It may be estimated with the standard $\epsilon$-expansion method.

The lowest order result in $\epsilon$ can be easily
obtained from the Operator Product Expansion (OPE)
coefficients\cite{Cardy}.
As a result, we obtain the scaling dimension $y_n$ of
$\lambda_n$ in $4 - \epsilon$ dimensions as
\begin{equation}
	y_n = 	4 - n +
	 \epsilon \left( \frac{n}{2} - 1 - \frac{n(n-1)}{10} \right)
		+ O(\epsilon^2).
\label{eq:yn1}
\end{equation}
$y_n$ is defined so that the effective strength
of the perturbation $\lambda_n (l)$ at scale $l$ is proportional
to $l^{y_n}$ near the XY fixed point.
The case $n=4$ is actually the special case $N=2$
of the ``cubic anisotropy'' on the 3D $O(N)$ fixed point\cite{Cardy}.
Extrapolating the $O(\epsilon)$ result to 3D ($\epsilon = 1$),
we see that the $Z_n$ perturbation is irrelevant at the
3D XY fixed point for $n \geq n_c$.
The threshold $n_c$ is estimated to be $4$ in $O(\epsilon)$
In fact, $n=2$ and $n=3$ corresponds to the 3D Ising
and 3-state (ferromagnetic) Potts model,
which do not belong to XY universality class.
Thus $n_c$ is expected to be at least $4$.
This is consistent with the above result from $O(\epsilon)$.
However, extrapolating the lowest order result in $\epsilon$ to
3D ($\epsilon = 1$) is not quite reliable; the true
value of $n_c$ might be larger than $4$.
On the other hand, we can make following observation.
For $n \geq 6$, $\lambda_n$ is marginal or irrelevant at the
3D Gaussian fixed point ($g=0$).
Thus it is natural to expect them to be irrelevant
at the more stable 3D XY fixed point, namely $n_c \leq 6$.
In fact, the numerical observation of
the 3D XY universality class in 6CL and AFP model
strongly suggests that $\lambda_6$ is irrelevant at the
XY fixed point and hence $n_c \leq 6$.
In the following, we restrict the discussion
to the irrelevant case $n \geq n_c$.

For the $O(2)$ symmetric case $\lambda_n=0$,
low-temperature phase $u < 0$ is renormalized to the
low-temperature fixed point.
It describes the massless Nambu-Goldstone (NG) modes on the
groundstate with the spontaneously broken $O(2)$ symmetry.
Let us call the low-temperature fixed point as NG fixed point.
In terms of the field theory, it is described by the
$O(2)$ sigma model (free massless boson field)
\begin{equation}
	S = \int d^3x \, \frac{K}{2} (\partial_{\mu} \phi)^2
\end{equation}
where $\phi$ is the angular variable $\Phi \sim |\Phi| e^{i \phi}$.
Namely, 
only the angular mode $\phi$ remains gapless as a NG boson.
In three dimensions, the coupling constant $K$
renormalizes proportional to the scale $l$, and goes to infinity
in the low-energy limit. The coupling constant may be absorbed
by using the rescaled field
$\theta = \sqrt{K} (\phi - \phi_0)$ so that the action is always
written as $\int d^3x \, (\partial_{\mu} \theta)^2 /2$.

Now let us consider effects of the symmetry breaking $\lambda_n$.
The symmetry breaking term can be written as
$- \lambda_n (\Phi^n + \bar{\Phi}^n) = - \lambda_n |\Phi|^n \cos{n \phi}$.
Using the rescaled field $\theta$, the total effective action
at scale $l$ becomes
\begin{equation}
S = \int d^3 x \frac{1}{2}  (\partial_{\mu} \theta)^2
- \lambda_n K^3 \int d^3 x \cos{[n (\phi_0 + \frac{\theta}{\sqrt{K}})]} ,
\end{equation}
where the factor $K^3 \sim l^3$ comes from the scale transformation
of the integration measure.
In the thermodynamic limit, we should take $K \rightarrow \infty$ limit.
Physically, it means that the $O(2)$ symmetry is spontaneously broken
so that the angle is fixed to some value $\phi_0$ in a single infinite
system. 
Then the Taylor expansion of the cosine in $\theta/\sqrt{K}$ becomes valid:
\begin{equation}
K^3 \cos{[n (\phi_0 + \frac{\theta}{\sqrt{K}})]} 
	= \sum_{j=0}^{\infty} c_j K^{3-j/2} \theta^j ,
\end{equation}
where
\begin{eqnarray}
c_{2k} &=& (-1)^k \frac{n^{2k}}{(2k)!} \cos{n \phi_0},  \nonumber \\
c_{2k+1} &=& - (-1)^k \frac{n^{2k+1}}{(2k+1)!} \sin{n \phi_0} ,
\end{eqnarray}
for a nonnegative integer $k$.
The five terms $j= 1, \ldots 5$ are relevant perturbations.
For any value of $\phi_0$, some of the coefficients $c_j$ of
these relevant terms are non-vanishing.
We therefore conclude that, unlike the 2D case,
the $Z_n$ perturbation is always relevant
for any value of $n$ at the NG fixed point.
We emphasize that this conclusion is universal in three dimensions
and independent of the microscopic model.
Shortly speaking, the $Z_n$ perturbation gives mass to
the pseudo NG boson $\theta$, which would be massless NG boson
in the absence of the perturbation.
In contrast, in two dimensions the coupling constant $K$ of the
free boson field theory is dimensionless,
and the above argument does not apply.
It is related to the absence of a spontaneous breaking of
a continuous symmetry.

\begin{figure}
\begin{center}
\epsfxsize=8cm
\epsfbox{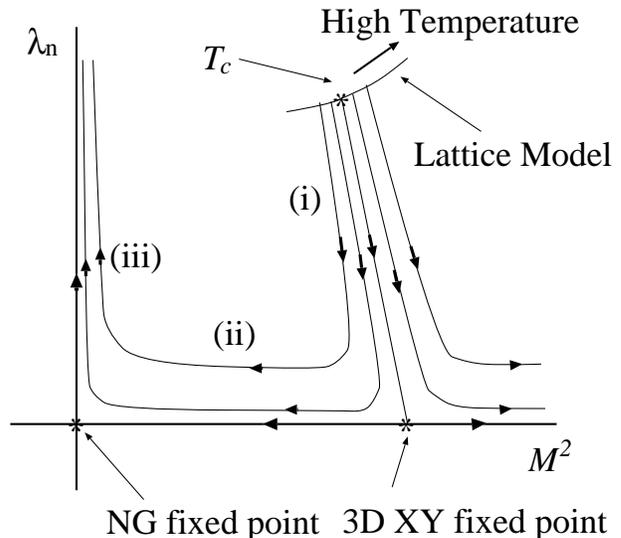}
\caption{The RG flow diagram of the $Z_n$ models, projected
onto the two-dimensional parameter space spanned by $u$ and $\lambda_n$.
The $Z_n$ perturbation $\lambda_n$ is irrelevant at the 3D XY fixed
point, but is relevant at the NG fixed point.
For $T$ slightly less than $T_c$, the RG flow is divided into
the three stages (i),(ii) and (iii).
}
\label{fig:RGflow}
\end{center}
\end{figure}

We now have a global picture of the RG flow as shown in
Fig.~\ref{fig:RGflow}.
The phase transition between the ordered phase and
the disordered phase is governed by the XY fixed point.
This means that the critical exponents are identical to
those of the XY model.
This is consistent with the numerical results.
In the disordered phase above $T_c$, there will be no essential effect
of the $Z_n$ perturbation.
However, the nature of the ordered phase is more interesting.
The $Z_n$ perturbation $\lambda_n$ is eventually enhanced
in the ordered phase below $T_c$.
It means that all region below $T_c$ belong to the
massive phase with the spontaneously broken $Z_n$ symmetry.
There is no rotationally symmetric intermediate phase,
unlike the 2D case.
Only a precisely $O(2)$ symmetric model with $\lambda_n =0$
is renormalized to the NG fixed point below $T_c$, corresponding to
the rotationally symmetric low-temperature phase.

An interesting aspect of the RG flow diagram is that
the $Z_n$ perturbation is irrelevant at the 3D XY fixed point
but is relevant at the low-temperature NG fixed point.
This could be related to a nontrivial system size dependence
found in a Monte Carlo Renormalization Group calculation\cite{KOMCRG}.
For $T$ slightly less than $T_c$, the symmetry breaking perturbation
$\lambda_n$ is renormalized to a small value by the RG flow,
and remains small until the RG flow reaches near the NG fixed point.
It means that the mass of the pseudo-NG bosons
is suppressed by the fluctuation effect.
At a finite scale (for example in a finite size system),
the ordered phase near $T_c$ is very similar to
the low-temperature phase of the XY model.
This naturally explains the numerical observation of the
apparently rotationally symmetric ``phase'' in 6CL
or the AFP model.
For larger $n$, the mass is more suppressed, and the 
low-temperature side of the transition appears to be $O(2)$ symmetric
until the system size becomes very large.
However, for any finite $n$,
the low-temperature side of the transition $T < T_c$
is not truly massless nor $O(2)$ symmetric in the thermodynamic limit,
as already pointed out.

\section{Scaling law in the ordered phase}

Based on the RG picture, we derive a scaling law on an
order parameter ${\cal O}_n$ which characterizes the symmetry breaking 
from the $O(2)$ to $Z_n$ symmetry.
There are various possible definitions of ${\cal O}_n$.
On the 6CL model,
Miyashita\cite{Miya} numerically measured
an order parameter $\Delta$ which 
corresponds to the effective barrier height.
On the AFP model, Heilmann, Wang and Swendsen\cite{HWS}
studied $\langle \phi_6 \rangle$, which is
the Fourier transform of the angle distribution density
of average spins.
The following consideration apply to the both cases.

\begin{figure}
\begin{center}
\epsfxsize=8cm
\epsfbox{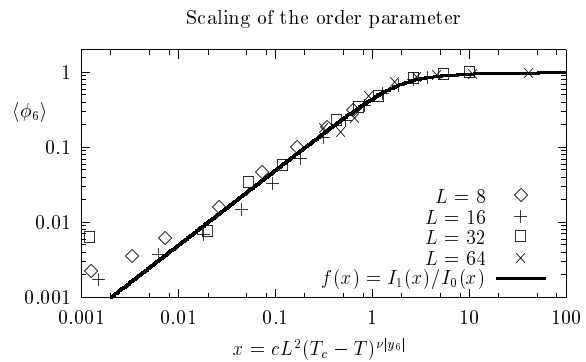}
\caption{The order parameter $\langle \phi_6 \rangle$ taken
from Ref.~\protect\cite{HWS}.
They are scaled by $x=c L^2 (T_c -T)^{\nu |y_6|}$, for
various system sizes and temperatures.
The data are consistent with the scaling law~(\protect\ref{eq:orderp})
with the exponent $\nu |y_6| = 4.8$.
They also agree with the approximate scaling function $f(x)=I_1(x)/I_0(x)$,
for $c=0.025$.}
\label{fig:scaling}
\end{center}
\end{figure}

For large enough $L$ and $T$ slightly lower than $T_c$
we divide the RG flow to three stages, as shown in Fig.~\ref{fig:scaling}
\begin{description}
\item[(i)] The RG flow near the 3D XY fixed point. The symmetry breaking
$\lambda_n$ is irrelevant, and
is renormalized proportional to $l^{- |y_n|}$ at length scale $l$.
\item[(ii)] The RG flow from the neighborhood of the  3D XY fixed point
to the NG fixed point. For simplicity, we assume that the
symmetry breaking $\lambda_n$ is unchanged in this stage.
\item[(iii)] The RG flow near the NG fixed point. 
$\lambda_n$ is relevant, giving a mass to the NG boson.
\end{description}
The length scale $l_c$, at which the crossover from Stage (i)
to (ii) occurs, is given by
$ l_c \sim \mbox{const.} (T_c -T)^{-\nu}$ ,
where $\nu$ is the correlation length exponent of the 3D XY
universality class. 
Thus, at the crossover,
\begin{equation}
\lambda_n \sim \mbox{const.} (T_c - T)^{\nu |y_n|}.
\label{eq:lmdscale}
\end{equation}
This also gives the effective value of the perturbation $\lambda_n$
at the crossover from Stage (ii) to Stage (iii).
 
In the presence of the $Z_n$ perturbation,
the spin configuration would be dominated by the ordered regions
which are separated by domain walls in a large system.
The free energy costed by the domain walls is proportional to
their area, which scales as $L^2$ for the system size $L$.
Therefore the effective
``barrier height'' is proportional\cite{Miya} to $L^2$.
Combining this with eq.~(\ref{eq:lmdscale}),  
we conclude that the order parameter is a function of
a single scaling variable:
\begin{equation}
	{\cal O}_n = f( c L^2 (T_c - T)^{\nu |y_n|}),
\label{eq:orderp}
\end{equation}
where $c$ is a constant.
The function $f$ is universal, but of course depends on the
definition of the ${\cal O}_n$.
While the scaling by $L^2$ was used in Ref.~\onlinecite{Miya}, we
find that the temperature dependence of the order parameter
is also governed by a scaling.
Interestingly, the exponent $\nu |y_n|$
is completely determined by the 3D XY fixed point.

\section{Comparison with the numerical results}

In the numerical study\cite{HWS} of the AFP model,
they claimed the existence
of the intermediate phase, in which the order parameter
$\langle \phi_6 \rangle$ is very small
even for relatively large lattice (upto $L = 64$).
However, we re-analyze their data to demonstrate
the scaling relation~(\ref{eq:orderp}),
and hence the validity of the RG picture.
In Fig.~\ref{fig:scaling}, we show the data
taken\cite{HWSremark} from Fig.~3 of Ref.~\onlinecite{HWS}.
We chose $\nu |y_6| = 4.8$ to give the best scaling.
The data for various temperature
and various system sizes fall remarkably
into a single curve as a function of
the scaling variable $x = c L^2 (T_c -T)^{\nu |y_n|}$.
This supports the proposed scaling relation~(\ref{eq:orderp}).
Furthermore, if we approximate the effective potential
by $ - x \cos{ 6 \phi}$, the scaling function is given by
\begin{equation}
f(x) = \frac{\int d \phi \, \cos{(6 \phi)} e^{x \cos{(6 \phi)}}}{
		\int d \phi \, e^{x \cos{(6 \phi)}}}
= \frac{I_1(x)}{I_0(x)} ,
\end{equation}
where $I_n$ is the modified Bessel function.
Choosing $c = 0.025$, the scaled data agree with this simple
function rather well.
We note that the data appear to deviate from the scaling law for small $x$.
This may be due to the insufficient system size $L$
or the relatively large statistical error.

We emphasize that the present scaling relation is a strong
evidence of the single phase transition at the temperature $T_c$.
In contrast, the scaling of the ``spontaneous magnetization''
$\rho = | \Phi | \propto (T_c -T)^{\beta}$
does not distinguish our picture and the ``intermediate phase''
scenario of Ref.~\onlinecite{HWS}.

On the other hand, 
the scaling function $f(x)$ for $\Delta$ in Ref.~\onlinecite{Miya}
is linear in $x$ by definition.
He indeed found that $\Delta$ is scaled by $L^2$.
However he did not discuss the temperature depedence.
We have attempted to analyze the data in Figs.~6 and 7 in
Ref.~\onlinecite{Miya}, to find that they are roughly consistent with our
scaling law~(\ref{eq:orderp}) with the exponent $\nu |y_6| \sim 4$.
The estimate is difficult because there are only small number
of temperature points available in Ref.~\onlinecite{Miya}.
According to our picture, the exponent $\nu |y_6|$ is a universal quantity
determined by the 3D XY universality class.
Considering the available data,
the above results on AFP and 6CL models
are consistent with the universality hypothesis, although not
conclusive.
It would be interesting to obtain more numerical data
on these models to check our scaling law~(\ref{eq:orderp}).

The exponent $\nu$ has been determined as $\nu \sim 0.67$
for the 3D XY universality class\cite{Zinn}.
Combining with the above estimates of $\nu |y_6|$,
$|y_6|$ is estimated as about $6$.
Unfortunately, the irrelevant eigenvalue $y_6$ has not been
much discussed in the literature.
The lowest order result~(\ref{eq:yn1})
in the $\epsilon$-expansion gives $|y_6| = 3$, which
is not quite consistent with the numerical estimate.
However, it is perhaps not surprising to obtain an inaccurate result
in the lowest order of the $\epsilon$-expansion.
It would be interesting to carry out the calculation to higher
orders in $\epsilon$, or to estimate $y_6$ by other means.

\section{Conclusion and Discussions}

In this article, we clarified a RG picture of phase structure
of 3D $Z_n$ symmetric models, which was introduced earlier by
Blankschtein et al.\cite{Blank}. 
There is no finite region of intermediate phase with
a (spontaneously broken) $O(2)$ symmetry, but only a crossover
to a massive phase where the discreteness of $Z_n$ is relevant.
Based on the RG picture, we have derived a scaling law of the
order parameter in the 3D $Z_n$ models
The existing Monte Carlo data\cite{HWS} on the AFP model,
which was used to claim the intermediate phase,
was shown to be consistent with the scaling law.
Thus we conclude that the RG picture is valid on the AFP model,
and there is only one transition at $T_c \sim 1.23$ with
the 3D XY universality class.

We would like to make a few final remarks.
Firstly, 
we note that the RG argument used in the present article does not
contradict to the transition of other than XY universality class,
because only the local stability of the XY fixed point was discussed.
It is possible that a lattice model with $Z_n$ symmetry
is renormalized to another (unknown) RG fixed point.
Actually, it appears somewhat controversial
whether the transition of the STI model belong to the
XY universality class\cite{PM,GBK}.
On the other hand, the available numerical results strongly
supports that the 6CL and AFP models at the critical temperature
are renormalized into the XY fixed point.
Once the transition is known to be XY universality class,
the RG picture and the scaling law discussed in this
article should apply to the ordered phase, for 
the temperature slightly below the critical point.

Secondly, as discussed in Ref.~\onlinecite{Blank},
the ``bare'' value of $\lambda_n$ (at a small length scale)
may have opposite sign in some circumstances.
Namely, the minima and maxima of the potential
of $\phi$ are swapped.
In such a case, the ordered phase may correspond to the
Permutationally Symmetric Sublattice
(PSS) phase proposed in Ref.~\onlinecite{RL}
for the AFP model, or the
Incompletely Ordered Phase (IOP) proposed in Ref.~\onlinecite{Mitsubo}
for the 6CL model. 
In the vicinity of the critical point, 
the temperature dependence of the bare $\lambda_n$
is not essential because
the leading dependence on the temperature is determined by
the critical effect, as shown in eq.~(\ref{eq:orderp}).
However, it may be important in a wider temperature range.
In particular, if the bare $\lambda_n$
changes sign at some temperature $T_L$ lower than $T_c$,
we would have a transition\cite{Blank} at $T_L$.
Such a transition would be controlled by the NG fixed point.
The existing numerical data indicates that there is no such phase
transition in the standard AFP model on the simple cubic lattice or
in the standard 6CL model.
However, such a transition may be possible
in some modified models\cite{Kasono,LR}.
In fact, Blankschtein et al.\cite{Blank} argued it to
exist in the STI model.

\acknowledgements

I would like to thank Hikaru Kawamura, Macoto Kikuchi, Ryo Kishi,
Seiji Miyashita and Yohtaro Ueno for useful discussions.
This work is supported in part by Grant-in-Aid for Scientific Research
from the Ministry of Education, Culture and Science of Japan.

\end{document}